\documentclass[%
 aapm,
 mph,%
 amsmath,amssymb,
 reprint,%
]{revtex4-2}

\usepackage{graphicx}
\usepackage{dcolumn}
\usepackage{bm}

\usepackage[mathlines]{lineno}

\usepackage{xcolor}
\usepackage{braket}
\usepackage{bm}
\usepackage{amsmath,amssymb}
\usepackage{graphicx}

\begin{document}

\title{Dynamic conditioning of two particle discrete-time quantum walks}

\author{Federico Pegoraro$^*$, Philip Held$^*$\footnote{These authors contributed equally}, Sonja Barkhofen, Benjamin Brecht, and Christine Silberhorn}

\affiliation{Integrated Quantum Optics Group, Institute for Photonic Quantum Systems (PhoQS),
	Paderborn University, Warburger Straße 100, 33098 Paderborn, Germany}
\vspace{10pt}

\begin{abstract}
In real photonic quantum systems losses are an unavoidable factor limiting the scalability to many modes and particles, restraining their application in fields as quantum information and communication. For this reason, a considerable amount of engineering effort has been taken in order to improve the quality of particle sources and system components. At the same time, data analysis and collection methods based on post-selection have been used to mitigate the effect of particle losses. This has allowed for investigating experimentally multi-particle evolutions where the observer lacks knowledge about the system's intermediate propagation states. Nonetheless, the fundamental question how losses affect the behaviour of the surviving subset of a multi-particle system has not been investigated so far. For this reason, here we study the impact of particle losses in a quantum walk of two photons reconstructing the output probability distributions for one photon conditioned on the loss of the other in a known mode and temporal step of our evolution network. We present the underlying theoretical scheme that we have devised in order to model controlled particle losses, we describe an experimental platform capable of implementing our theory in a time multiplexing encoding. In the end we show how localized particle losses change the output distributions without altering their asymptotic spreading properties. Finally we devise a quantum civilization problem, a two walker generalisation of single particle recurrence processes.
\\
\\
\textbf{E-mail:} \verb|federico.pegoraro@upb.de|\\\\
$^*$These authors contributed equally to this work

\end{abstract}

\maketitle
%
%
%
%

%

\bibliographystyle{unsrt}

\section{Introduction}
Quantum walks, the quantum analogy of the classical random walks, have been established as a variable test bed for complex quantum algorithms on large networks \cite{aharonov1993quantum,kempe2003quantum, venegas2012quantum}.
Analogously to its classical counterpart, the dynamics of a discrete-time quantum walk (DTQW) consists of a quantum coin toss, which determines the direction of the subsequent conditional step operation.
DTQWs were applied to the simulation of a great variety of topics, e.g., topological effects \cite{nitsche2019eigenvalue, barkhofen2017measuring, barkhofen2018supersymmetric}, recurrence \cite{nitsche2018probing}, percolation \cite{elster2015quantum}, higher-dimensional graphs \cite{schreiber_2d_2012, lorz2018photonic, xue2015localized,de2020experimental} and disorder \cite{schreiber_decoherence_2011, bagrets2021probing}. Experimental implementations of quantum walks involve nuclear magnetic resonances \cite{du2003experimental, ryan2005experimental}, trapped ions \cite{schmitz2009quantum,zahringer2010realization}, atoms \cite{karski2009quantum, genske2013electric}, but most commonly photonic platforms based on beam splitter (BS) cascades \cite{bouwmeester1999optical, do2005experimental}, beam displacers \cite{kitagawa2012observation,broome2013photonic, xue_experimental_2015} waveguides \cite{perets2008realization, bromberg2009quantum, peruzzo2010quantum,owens2011two, sansoni_two-particle_2012, di2013einstein, crespi2013anderson, meinecke2013coherent, poulios2014quantum}, fibre loops \cite{schreiber_photons_2010,regensburger2011photon, schreiber_2d_2012, barkhofen2017measuring,lin2022observation}, photonic chips \cite{tang2018experimental}, multimode fibres \cite{defienne2016two} or the adoption of the orbital angular momentum space of light \cite{cardano_quantum_2015}, see \cite{wang2013physical} for a review of the physical implementations of quantum walks.

In most experimental works, only one quantum particle evolves coherently in a discrete network.
Yet, the introduction of multiple quantum particles to passive networks is essential in many quantum communication and computation schemes \cite{knill2001scheme, childs2009universal, rohde2011multi}.
Only when more than one walker is involved in the quantum walk dynamics, collective quantum effects, such as many-particle correlations can be studied \cite{stefanak2011directional}.
Only  a few experimental implementations of DTQWs with more than one walker have been demonstrated, e.g. to study  bosonic and fermionic behaviour \cite{ sansoni_two-particle_2012}, disorder \cite{crespi2013anderson, laneve2021enhancing}, the interplay of first and second order coherences \cite{nitsche2020local}, mimicking two-dimensional lattices with one walker \cite{xue2015localized} and two walkers in two-dimensional lattices  \cite{esposito2022quantum}.

The strongest effect that prevents the scaling to high numbers of walkers in the experiments is the inevitable loss of particles due to non-ideal efficiencies of the network components and the measurement devices. Since the network efficiency $p$ scales exponentially with the number of travelling particles $N$, we find a very unfavourable scaling for multi-photon quantum walks. In addition, we must keep in mind the mainly probabilistic nature of quantum states sources.
In order to overcome losses, technological advances, such as in the engineering of better and more efficient quantum sources and detectors and almost loss-free optical components are pushed forward, aiming at on-demand photon sources and ideal network properties. In addition, post-selection on coincidence events involving the targeted number of walkers provides a valid selection criterion for a successful measurement run, in which all initiated walkers have arrived at their final step.

Nevertheless, under certain conditions the introduction of controlled losses in a network is even indispensable to model physical effects, e.g. recurrence phenomena \cite{stefanak_recurrence_2008,stefanak2008recurrence,nitsche2018probing}, non-hermitian evolutions and parity-time symmetry \cite{wimmer2015observation,mochizuki2016explicit,zhan2017detecting,kremer2019demonstration,xia2021nonlinear,xiao2021observation}, or invasiveness of quantum measurements \cite{smirne2019experimental} just to name a few. Furthermore, the adoption of this kind of scheme is necessary when addressing the dynamics of an open quantum system. 
For these reasons, the main objective of this work is to study the fundamental effect of mode dependent particle losses in a DTQW of two indistinguishable photons.

In order to do this, we have devised a method to perform a quantum walk of $n$ temporal steps where we can insert losses at a time step $m<n$ on a selected propagation mode in a controlled way and successively detect the output single particle distributions resulting from the remaining unperturbed dynamic. We will first describe this from a theoretical point of view, in particular, we will illustrate the theoretical formalism developed to treat the problem of losses by coupling the system with an external ancillary mode and conditioning on events where we have only one particle in a system output mode. For this reason we will refer to this method as \emph{dynamic conditioning}.
We will show how the dynamic conditioning can be implemented in a time-multiplexing (TM) architecture, where time encoding is used to realise the system evolution. In this architecture we are able to access intermediate stages of the evolution to insert losses in a known mode and we will show how this knowledge change the single photon output with respect to the unperturbed case. We will then see, how cancelling this information by averaging on all the possible loss configurations the distributions recover the features of the ones expected from a lossless system. In the end we will analyse the spreading of the conditioned distributions. From this we will see that the dynamic conditioning does not change the ballistic behaviour typical for quantum walks. In conclusion, our dynamically conditioned quantum walk constitutes a prime example of non-unitary evolution and poses the basis for the study of more complex systems and problems, like a quantum civilization problem that we will describe in the final part of this work.

The paper is structured as follows:
In Sec.~\ref{Sec:theory} we explain the theoretical modelling of the two-particle quantum walk and the dynamic conditioning.
This is followed by the introduction of the experimental platform in Sec.~\ref{Sec:platforms}.
The results of the measurements and the dynamic conditioning are presented in Sec.~\ref{Sec:results}.
We conclude the article in Sec.~\ref{Sec:conclusion} and give an outlook on future research directions.

\section{Theory Background}
\label{Sec:theory}
DTQWs describe the discrete time evolution of one or many walkers on a grid of separate vertices.
For a one-dimensional lattice the single walker state is defined as
\begin{equation} 
|\Psi (t)\rangle = \sum_{\substack{x\in \mathbb{Z}, \\ c \in\{H,V\}}} \Psi_{x,c}(t)|x,c\rangle \in \mathcal{H}_x \otimes \mathcal{H}_c
\end{equation}
with the discrete positions states $\{\ket{x}\}_{x\in \mathbb{Z}}$ and the two-dimensional coin states $\{\ket{c}\}_{c=H,V}$ belonging to the Hilbert spaces $\mathcal{H}_x$ and $\mathcal{H}_c$.
$|\Psi_{x,c}(t)|^2$ indicates the probability of finding the walker at position $x$ in coin state $c$ at step $t$.
For all $t$, $  \sum_{x, c}|\Psi_{x,c}(t)|^2 = 1$ ensures normalised probability distributions.

In order to illustrate the evolution of a DTQW we identify the two coin base states with the vectors: $\ket{H}=(1,0)^T$ and $\ket{V}=(0,1)^T$.
 The walker's state changes over a time step via the application of coin operation, which is potentially time and position dependent,
\begin{equation}
	\label{eq:coin}
\hat{C}(\varphi_{x,t}) = \sum_x |x\rangle \langle x| \otimes \begin{pmatrix}
\cos(\varphi_{x,t}) &\sin(\varphi_{x,t})\\
\sin(\varphi_{x,t}) & -\cos(\varphi_{x,t})
\end{pmatrix}
\end{equation}
and rotates the coin state at position $x$ and time $t$ in the H/V basis of an angle that depends on the phase parameter $\varphi_{x,t}$ .
Note, that the phases used in the coin operator (\ref{eq:coin}) may differ depending on which optical element realises the rotation.
The convention used here corresponds to the Jones matrix of a half-wave plate.
Throughout this work we adopt the Hadamard coin, i.e. $\varphi_{x,t} = 45^\circ$ $\forall$ $x$ and $t$, which results in a 50:50 splitting at each position in each time step, defined as 
\begin{equation}
\hat{C}_H= \hat{1}_x \otimes\frac{1}{\sqrt{2}} \begin{pmatrix}
1 &1\\
1 & -1
\end{pmatrix} ~~.
\end{equation}

This quantum coin toss is followed by a conditional position shift, performed by the step operator, defined as
\begin{equation} 
\begin{aligned}
\hat{S} = &\sum_{x}
\big( \ket{x+1}\!\bra{x}\otimes \ket{H}\!\bra{H}
+ \ket{x-1}\!\bra{x}\otimes \ket{V}\!\bra{V} \big)~~.
\end{aligned}
\label{eq:stepoperator}
\end{equation}
The two together form one step of the unitary walk evolution  governed by $\hat{U} = \hat{S}\hat{C}_H$.

The spatial probability distribution of a walk initialised with a single walker input state $|\Psi_\mathrm{in}\rangle$ evolved over $N$ steps derives form the action of $\hat{U}$ applied $N$ times on $\ket{\Psi_{in}}$ and can be calculated as
\begin{equation}
P(x) = \mathrm{Tr}[(\ket{x}\bra{x} \otimes 1_C) \hat{U}^N |\Psi_\mathrm{in}\rangle \langle\Psi_\mathrm{in}| \hat{U}^{N\dagger}]~~.
\end{equation}
When the measurement unit additionally provides coin resolution, we obtain the following probability distribution
\begin{equation}
P(x,c) = \mathrm{Tr}[(\ket{x}\bra{x} \otimes\ket{c}\bra{c}) \hat{U}^N |\Psi_\mathrm{in}\rangle \langle\Psi_\mathrm{in}| \hat{U}^{N\dagger}]~~.
\end{equation}

In the next step we model a two-photon walk initialised with two indistinguishable photons that obey bosonic particle exchange symmetry.
The two walker state lives in the global symmetric Hilbert space $\mathcal{H}_{1,2}^{\mathrm{sym}} =  \mathcal{H}_1 \otimes \mathcal{H}_2$, with the single walker Hilbert space $\mathcal{H}_{i} =  \mathcal{H}_{i,x} \otimes \mathcal{H}_{i,c}$.
Since we assume non-interacting walkers, the walk evolution is simply governed by the tensor product of two single particle walk unitarys $\hat{U}_{i}$
\begin{equation}
\hat{\mathcal{U}}_{1,2} = \hat{U}_1 \otimes \hat{U}_2~~.
\end{equation}

Two single-particle states $ |\Psi_1\rangle$ and $ |\Psi_2\rangle$ have to be combined in a symmetric way to correctly describe the bosonic statistics of the photons, thus the global state reads
\begin{equation}
 |\Psi\rangle_{1,2}^\mathrm{sym} = \frac{ |\Psi_1\rangle_1\otimes |\Psi_2\rangle_2 + |\Psi_2\rangle_1\otimes |\Psi_1\rangle_2}{\sqrt{2(1+|\langle \Psi_1|\Psi_2\rangle|^2)}}~~.
\end{equation}
In the demonstrated experiments based on polarisation photon pairs entering at the central node $ x =0$ in the network the photonic initial state looks like 
\begin{equation} \label{eq:initialstate}
	|\Psi\rangle_\mathrm{in} = \frac{ |0,H\rangle_1\otimes |0,V\rangle_2 + |0,V\rangle_1\otimes |0,H\rangle_2}{\sqrt{2}}~~.
\end{equation}

In contrast, classically indistinguishable particles are just modelled by the tensor product without symmetrisation
 \begin{equation}
 |\Psi\rangle_{1,2}^\mathrm{cl} = |\Psi_1\rangle_1\otimes |\Psi_2\rangle_2~~.
 \end{equation}
 In order to simplify the notation the index identifying particles will be dropped most of the time,
 when there is no possibility of confusion.
 In the classical setting also the measurement projector has to be symmetrised
 \begin{equation}
\hat{\Pi}_{i,j}^\mathrm{cl} =\frac{ |i,j\rangle\otimes \langle i,j|+|j,i\rangle\otimes \langle j,i|}{2}~~,
\end{equation}
where the index $i$ denotes a multi-index of coin and position mode.

In this work we study the conditioned statistics of two photonic walkers, which means that one of the photons is measured at a certain time and position, while the other continues its propagation. 
The remaining photon continues its evolution governed by $\hat{U}$ until it is also measured in the final step and the walk stops.
Since the walkers are indistinguishable, the partial projection operator, that mimics the measurement of one of the photons in mode $(x,c)$, also must obey the bosonic symmetry.
We will model this operation as
 \begin{align} \nonumber
\hat{\mathcal{M}}_{x,c} &=|x,c\rangle \langle x,c|\langle x,c|+ \\ \label{eq:proj}
 &\sum_{(x^\prime,c^\prime) \ne (x,c)} |x^\prime,c^\prime\rangle \frac{\langle x,c|\langle x^\prime,c^\prime| +\langle x^\prime,c^\prime|\langle x,c|}{\sqrt{2}}
 \end{align}

which acts on the two photon state mapping the contribution of the (symmetric) two photon wavefunction relative to mode  $(x,c)$ onto the corresponding single photon component.
This operator is symmetric under particle exchange and incorporates the indistinguishability of the two evolving photons.

After the partial projection in step $M$ and mode $(x,c)$, the remaining walker evolves further according to the single particle evolution $U_1$ until it arrives at step $N$ where it is measured.
Finally, its state looks like 
 \begin{equation}\label{eq:Psi_cond}
|\Psi\rangle_1 =\hat{U}_1 ^{N-M} \hat{\mathcal{M}}_{x,c}\hat{\mathcal{U}}_{1,2}^M| \Psi\rangle_{1,2, \mathrm{init}}^\mathrm{sym}~~,
\end{equation}
from which one can directly obtain the single particle probability distribution $P(x,c)$.

This approach is easily to generalise to larger multi-particle systems and evolutions, effectively providing a method to treat the dynamics of the open system constituted by the subset of surviving particles without having to resort to methods based on the solution of master equations such as Redfield \cite{redfield1965theory},  Gorini–Kossakowski–Sudarshan–Lindblad \cite{breuer2002theory}, Caldeira-Lagget \cite{caldeira1981influence} or Nakajima-Zwanzig \cite{nakajima1958quantum,zwanzig1960ensemble} equations, whose solution could be, in general, of nontrivial determination.

In order to simulate the loss of a photon in an indeterminate mode we need to average over all possible quantum walk patterns involving this mode
\begin{equation}
\bar{P}(x,c) = \sum_{k=1}^K a_k P_k(x,c) 
\end{equation}
where the weights $a_k$ sum up to 1 and the sum runs over the output distributions of the associated quantum walk distributions.
In the result section we will consider the case in which we suppose that one photon can be lost in any mode with equal probability setting $a_k = 1/K~\forall k$, with $K$ equal to the number of possible modes.

To quantify the agreement between sets of theoretically expected and experimentally estimated distributions, $P^\mathrm{theo}_{i}$ and $P^\mathrm{exp}_{i}$ we use the \emph{similarity} defined as: 
\begin{equation}
\mathcal{S} = \frac{\sum_{i} P^\mathrm{theo}_{i}\cdot P^\mathrm{exp}_{i}}{\sqrt{\sum_{i} (P^\mathrm{theo}_{i})^2\cdot\sum_{i} (P^\mathrm{exp}_{i})^2}}~,
\label{eq:simi}
\end{equation}
where $i$ is in general a multi-index that could include combinations of step, position and coin state of the walk. As a general remark, since we will consider normalised non negative quantities, $\mathcal{S}$ takes values in the interval [0,1], where a similarity of 0 implies perfectly decorrelated sets and a similarity of 1 can only be achieved when comparing two identical sets.

\section{Experimental Platform}
\label{Sec:platforms}
Photonic quantum networks and quantum walks can be found in various implementations, for a review of the experimental platforms see \cite{wang2013physical}. A suitable platform to implement the dynamic conditioning modelled in eq.~(\ref{eq:Psi_cond}) must necessarily provide access to intermittent stages of the walk dynamics. 
This means that one of the photons can be measured in an arbitrary position and coin mode at any time step without destroying the remaining one.

For a DTQW this is represented schematically in Fig. \ref{Sec:platforms}(a). The quantum walk evolution is realised by a BS cascade (light blue bars). At each step $N$ the blue (red) paths represent the $\ket{x,c}$ modes, with $c=H$ ($c=V$) and $x$ equal to an integer with the same parity of $N$ from the interval $[-N,N]$. The conditioning can be realised by coupling the system to an ancillary mode $\ket{anc}$ (represented in green in the figure), this is achieved by inserting an additional 50:50 BS (brown vertical bar) on the conditioning mode $\ket{x',c'}$ to probabilistically send only one particle to the ancillary mode. This results in the action of the operator: 
\begin{equation}
\hat{\mathcal{P}} = \sum_{(x^\prime,c^\prime)} |x^\prime,c^\prime\rangle \frac{\langle \mathrm{anc}|\langle x^\prime,c^\prime| +\langle x^\prime,c^\prime|\langle \mathrm{anc}|}{\sqrt{2}}~,
\end{equation}
on the two photon state. It is easy to see that the combination of this operation with the postselection on coincidence events involving only the ancillary and quantum walk modes of a later step results in the same operator as outlined in the theory section, see eq.~(\ref{eq:proj}).
Thus, these coincidence counts witness the dynamic conditioning modelled in eq.~(\ref{eq:Psi_cond}).

In this section, we will first describe the photon pair source used to produce the two walkers and then we will present a setup, consisting of a looped Mach-Zehnder interferometer that exploits TM to realise the lossy evolution.  

\begin{figure*}[t]
	\centering 
	\includegraphics[width =0.85\textwidth]{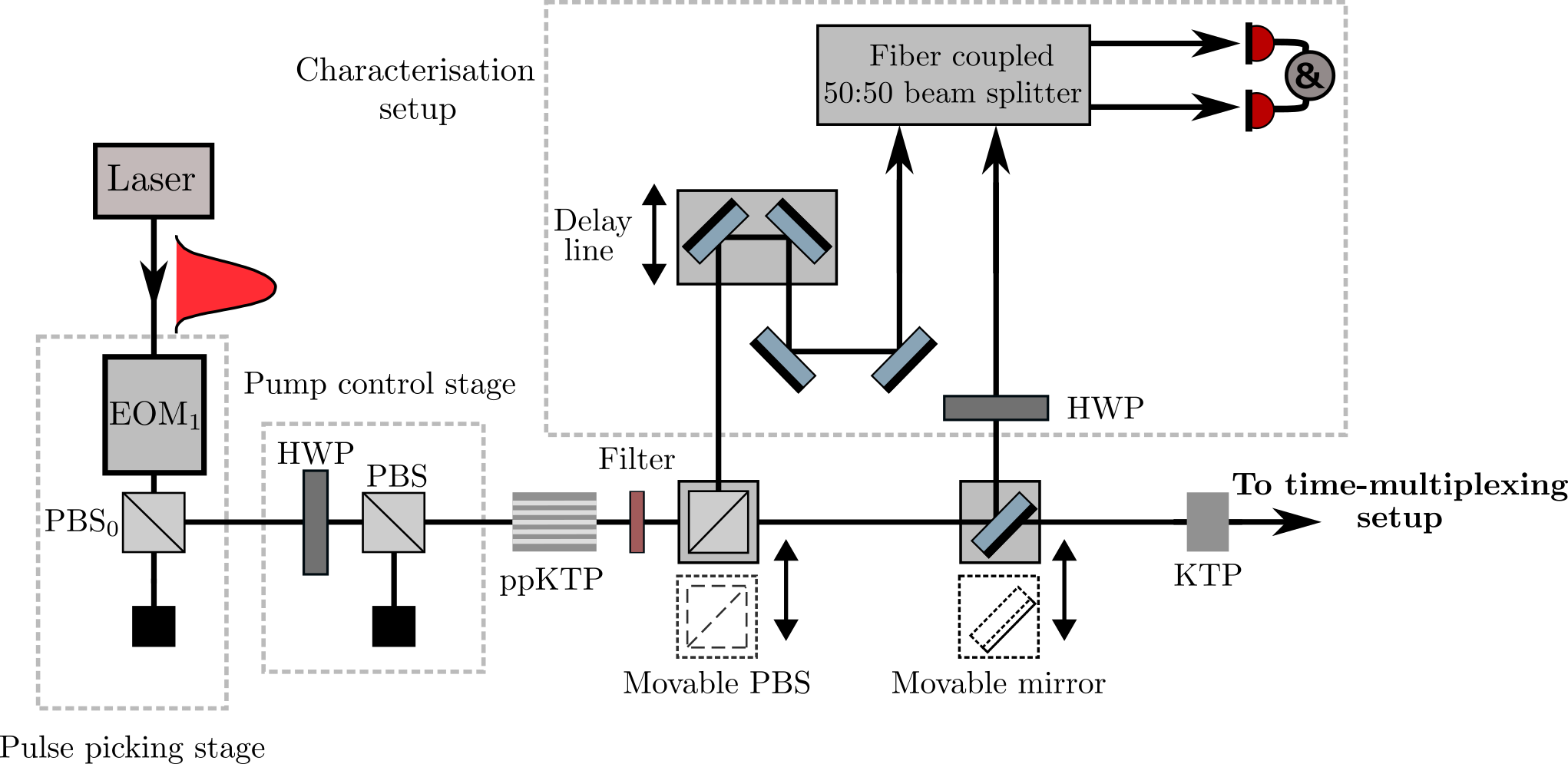}
	\caption{The photon pairs are produced in a periodically poled potassium titanyl phosphate (ppKTP) waveguide and spectrally filtered. The second KTP  crystal compensates the walk-off between the signal and the idler photon in the type II process when we sent them to the TM setup (Subsec. \ref{sub:TML}). EOM$_1$ picks the pump pulses and controls the repetition rate of the experiment. A movable PBS and mirror can be used to direct the photon pair to the characterisation setup. A delay line and half-wave plate are used to match the signal and idler arrival times and polarisation, before interfering them on a 50:50 fiber BS .}\label{Fig:source}
\end{figure*}

\begin{figure}[t]
	\centering 
	\includegraphics[width =\columnwidth]{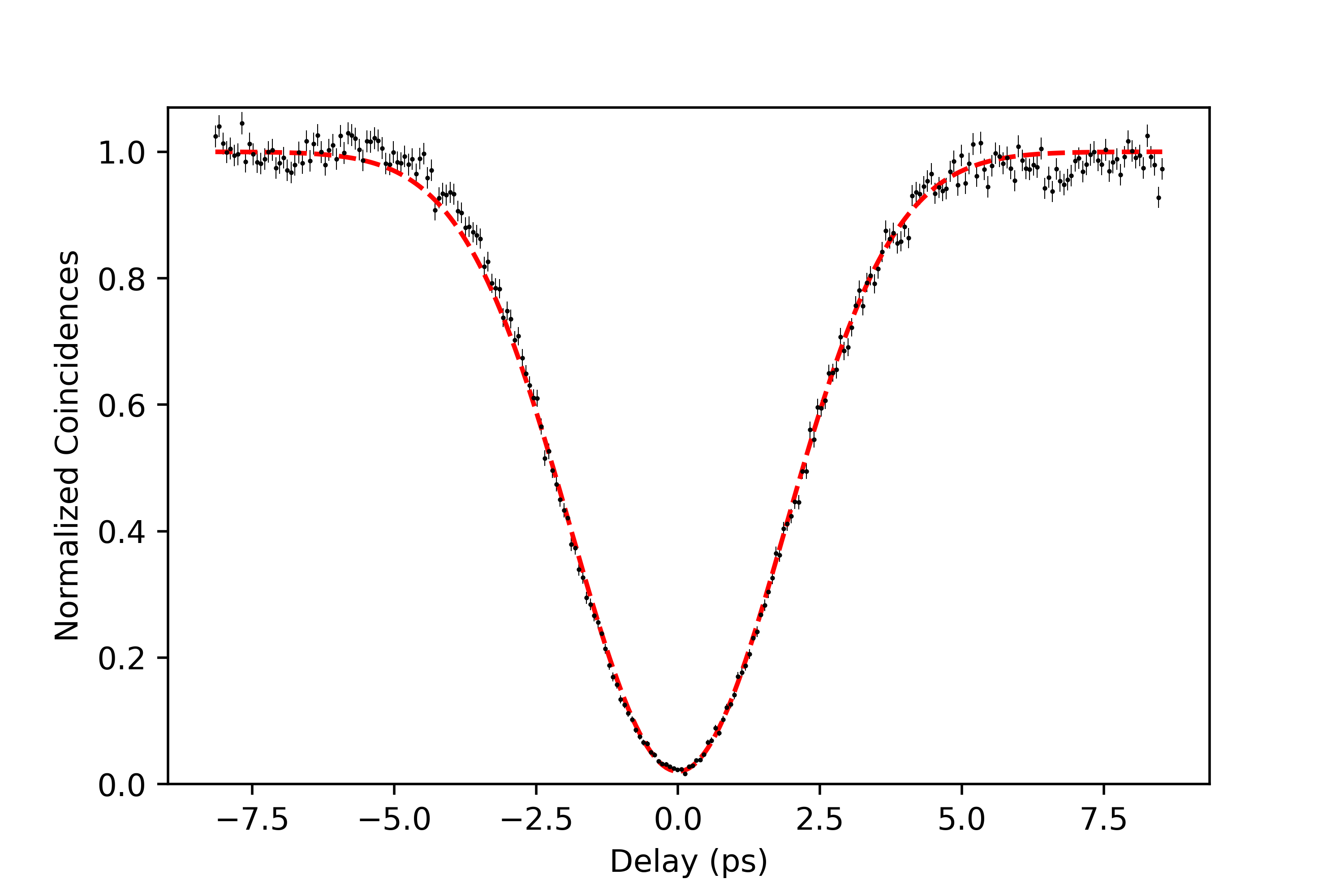}
	
	\caption{Hong-Ou-Mandel interference \cite{hong_measurement_1987} of the signal and idler photon generated in the type II PDC source and interfered at a 50:50 BS. The gaussian fit $1-0.980\exp(-0.139t^2)$ yields a raw visibility of 98.0\%. 
	}\label{Fig:HOM}
\end{figure}

\subsection{Photon pair source}
\label{Sub:source}
The two walkers are produced using a type-II parametric down-conversion (PDC) source based on a periodically poled potassium titanyl phosphate (ppKTP) waveguide. This process is capable of generating pairs of signal-idler photons that are orthogonally polarised and have high spatial and spectral purity \cite{harder2013optimized, nitsche2020local}.

Fig. \ref{Fig:source} shows a schematic drawing of the source with its components. We pump the PDC process using a laser that produces pulses with duration of the order of picoseconds at 772.5\,nm, with a bandwidth of $\approx 0.3$\,nm and a repetition rate of 76.4\,MHz. The pump laser passes through the electro-optic modulator $EOM_1$ (see Fig. \ref{Fig:source}) that is used in combination with the polarising BS $PBS_0$ to perform pulse picking and reduce the repetition rate. After this stage, a half-wave plate and another PBS are used to adjust the power of the pump before coupling it into the 2.5\,cm long ppKTP waveguide. As we want to operate the source in a regime where only two photons at a time are generated, we set the pump power at a level resulting in a generation probability of $p=10^{-1}$.
The waveguide generates relatively long ($\approx 3.2 $\,ps) photon pulses at a wavelength of $\approx1550$\,nm. The pulses are then filtered using a spectral filter with a central wavelength of 1545.22\,nm and a bandwidth of 1.8\,nm. A PBS and a mirror mounted on two manual translation stages can be moved into the beam path to send the photons either to a characterisation setup or to the TM setup (see Subsec. \ref{sub:TML}).

In the first setting, the reflected photons pass through a motorised delay line, while the transmitted ones encounter the movable mirror and an half-wave plate. Using the delay line and the wave plate we are able to synchronise the pulse pairs and match their polarisation to make them indistinguishable. Instead, if the PBS and mirror are moved out, the partner photons take the same optical path and, before entering the TM setup, they find a KTP sample that compensates for the polarization induced walk-off due to the waveguide birefringence.

We estimate the performance of the source with its Klyshko efficiency \cite{klyshko1980use}, defined as the ratio $r_c/r_s$ where $r_c$ is the rate of coincidence counts between signal and idler and $r_s$ is the single count rate of signal photons. With this source, we achieve a Klyshko efficiency of 36\% from PDC directly to the detectors. This parameter quantifies the overall efficiency of the source and effectively represent the probability for a generated photon pair to be transmitted through the setup and eventually produce a detection event.

\begin{figure*}[t]
	\centering 
	\includegraphics[width =\textwidth]{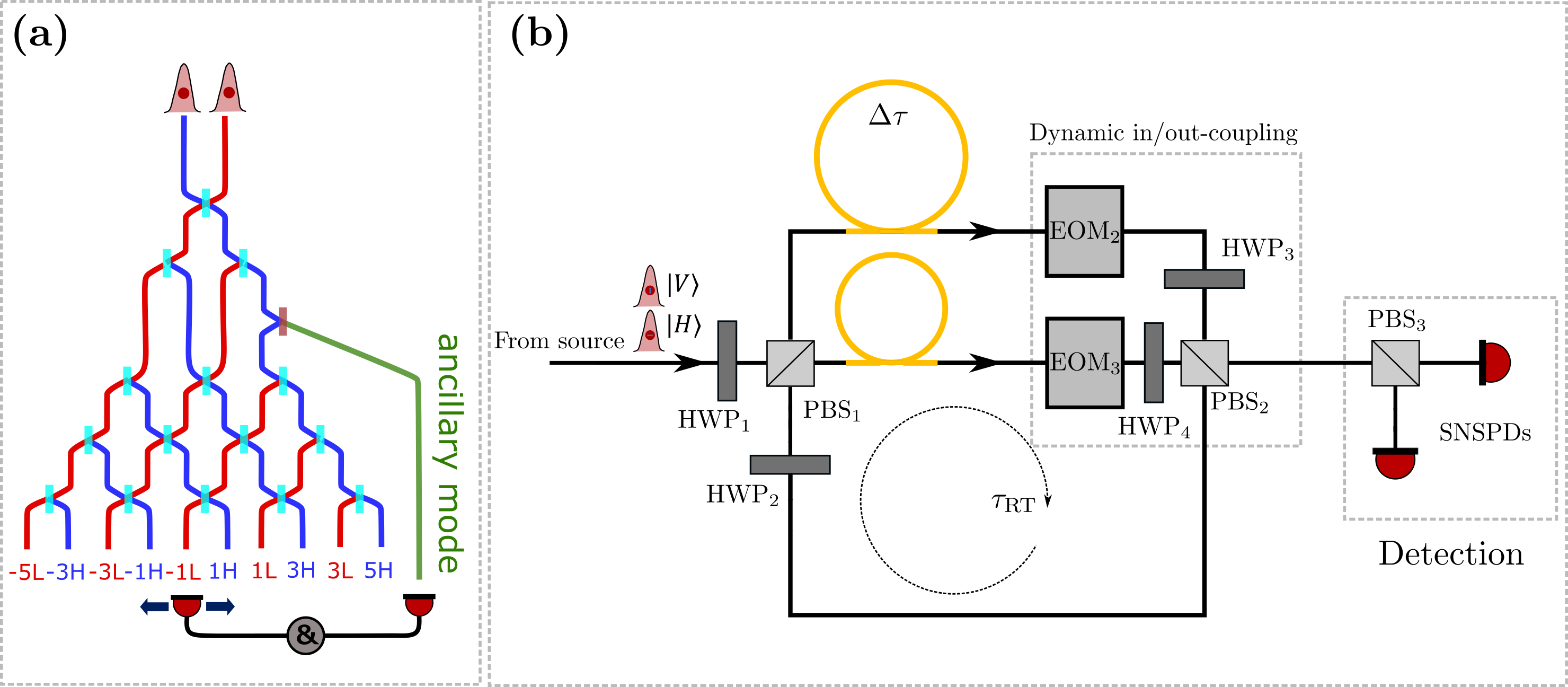}
	\caption{\textbf{(a) Schematic representation of the dynamic conditioning:} the input photons are produced in the type II PDC shown in Fig. \ref{Fig:source}. The turquoise bars represent the  BSs with a 50:50 splitting implementing the coin (see eq.~(\ref{eq:coin})) and the brown bar denotes the controlled loss by a single 50:50 BS, here exemplary shown for mode $(2,H)$ in step 2. Coincidences between the detector in the ancillary mode and one of the quantum walk modes signify a valid event of the dynamic conditioning of eq.~(\ref{eq:Psi_cond}).\\\textbf{ (b) TM setup:} the photons are deterministically coupled in using EOM$_2$ and EOM$_3$. In each round trip the HWP$_3$ and HWP$_4$ couple out a small portion of the light (15\%), which will be routed to the polarisation-resolved detection stage containing two SNSPDs. The loop photons pass HWP$_2$ which acts as the coin operator and are split afterwards according to their polarisation. Single-mode fibres of different length introduce a well-defined time delay $\Delta \tau$ between the constituents, which completes the translation of the spatial degree of freedom into the temporal domain.}
     \label{Fig:setups}
\end{figure*}

In order to quantify the indistinguishability of the produced photons we use Hong-Ou-Mandel (HOM) interference \cite{hong_measurement_1987}. To do this, we send the two photons to the BS and we record the coincidences between the two output ports while scanning the relative delay using the motorised stage, see Fig.~\ref{Fig:source}. As a measure of the source quality, we evaluate the raw visibility of the coincidence count suppression between signal and idler photon of up to $V = 98.0\%$, see Fig.~\ref{Fig:HOM}, demonstrating a very high indistinguishability between the partner photons.

\subsection{Time-multiplexing setup}\label{sub:TML}
The experimental setup is implemented by a time multiplexing architecture.
This established platform maps the position degree of freedom onto the time domain by introducing well defined time delays between the pulses \cite{schreiber_photons_2010,schreiber_decoherence_2011}.
At the core of this experimental setup lies a fiber-based unbalanced Mach-Zehnder interferometer with a feedback loop, as sketched in Fig. \ref{Fig:setups}(b).
The walkers are realised by two photons produced using the source described in Subsec. {\ref{Sub:source}}. In this implementation, the polarisation of the two particles acts as the coin degree of freedom.

The first coin operation eq.~(\ref{eq:coin}) is accomplished before PBS$_1$ by the half-wave plate HWP$_1$, set at $22.5^{\circ}$ (see Fig.{\ref{Fig:setups}}). It rotates the polarisation of the propagating photons and the following PBS$_1$ splits them up accordingly. Two single-mode fibres of 1085\,m and 1120\,m length in the arms of the interferometer lead to a position separation of 171.6\,ns and a roundtrip time of  $5.3227\,\mu$s, which constitutes the step separation. Both timings are matched to the parameters of the setup.
The conditional routing of the photons through the long or the short fibre realises the (temporal) shift in the  step operation according to eq.~(\ref{eq:stepoperator}).

After PBS$_2$ the pulses are fed back into the loop, where they pass through HWP$_2$ that performs the next coin operation and the dynamics continue.
Each time bin in the resulting pulse train can be uniquely associated to a step number and position. In the arms of the interferometer, two fast electro-optic modulators (EOMs), capable of implementing polarisation flips for individual time-bins, are used to realise the deterministic incoupling of the photons as already done in \cite{nitsche2018probing, nitsche2020local}. In this work, they are complemented by two half-wave plates at $11.4^\circ$  introducing a probabilistic 85:15 splitting at PBS$_2$. This means that in each step, the photons are routed directly to the detection with a probability of 15\%, while they will stay in the loop with 85\% probability.

This ratio provides the ideal trade-off between the reduction of the roundtrip losses to observe many quantum walk steps on the one hand and the possibility of having events  including artificial photon loss. In fact this allows to reconstruct the conditioned probabilities by recording data over an extended period of time and then post-selecting on events where the first photon was detected in an earlier step and the other measured in a later one.

\begin{figure*}[t]
	\centering 
	\includegraphics[width =0.9\textwidth]{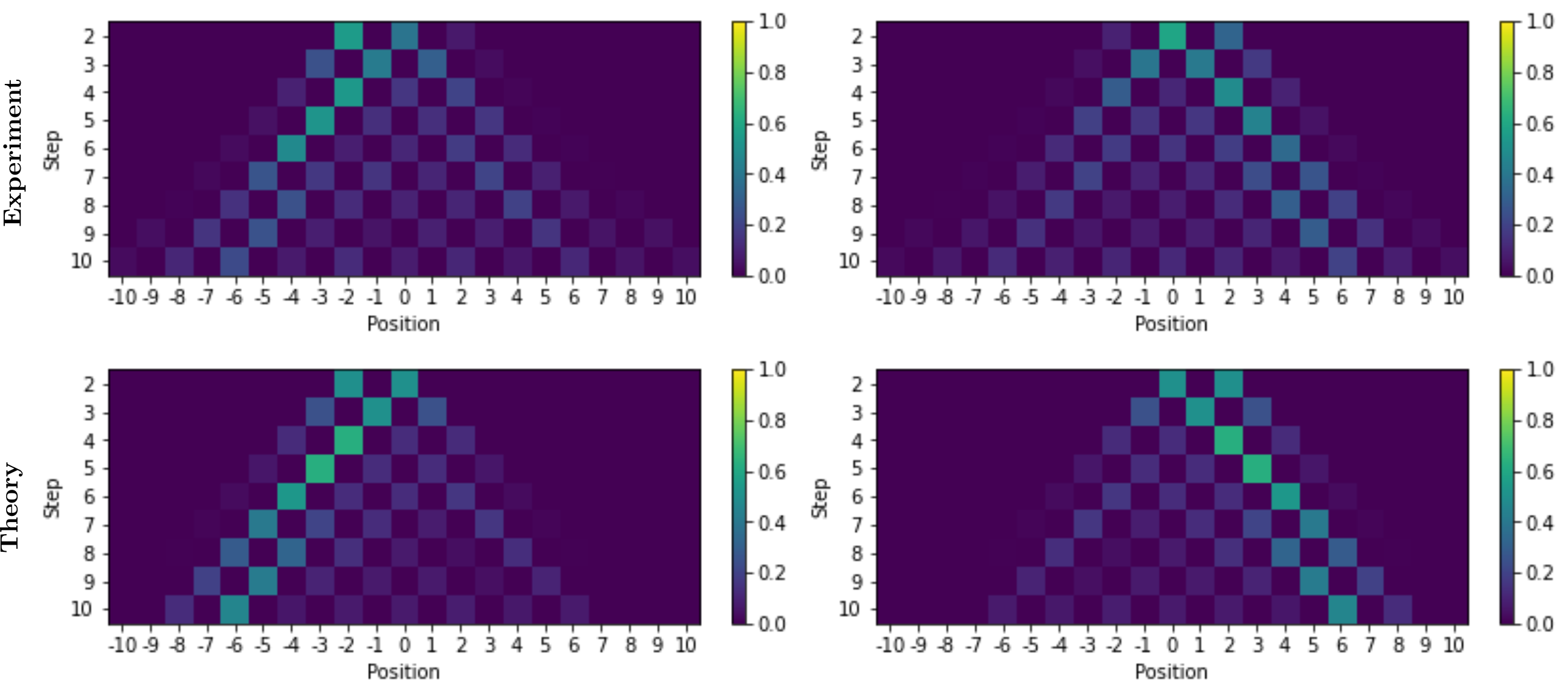}
	\caption{Output probability distributions for the dynamic conditioning at step 1. The pictures correspond to the conditioning modes $(V,-1)$ (left column) and $(H,1)$ (right column) in the experiment (upper) and theory (lower row). As aresult of the initial HOM interference we see how the peaks of the position probabilities stay localised on one side of the line. The mean similarities between experimental probability distribution and theoretical prediction (averaged over steps 2-10) are 96.05\% and 93.87\%, respectively.
	}\label{Fig:condstep1}
\end{figure*}

The pulse picking stage described in Subsec. \ref{Sub:source}, enables the reduction of the repetition rate of the pump laser of 76.4\,MHz to experimental repetition rates of the order of 10\,kHz; this guarantees that all roundtrips of one quantum walk are completed before the next run starts. Due to the common path geometry in the interferometer the photon pulses maintain a good indistinguishability even after several round-trips through the network.
The compatibility of the generated photons and the fibre network was already demonstrated in the context of local and global Hong-Ou-Mandel interference effects \cite{nitsche2020local}.

In the detection unit of the TM setup the light is split up according to its polarisation at PBS$_3$. Each of the two output ports is connected to one superconducting nanowire single photon detector (SNSPD) with efficiencies $\approx80\%$ and dead time of $\approx$70\,ns, which is well below the time-bin separation, allowing both polarization and time-bin resolved measurements. This setup shows a Klyshko efficiency of $20\%$ when considering two photons that enter it at PBS$_1$, then travel through the two fiber arms and exit the setup to be detected after PBS$_3$.

\section{Experimental Results}
\label{Sec:results}

\emph{\textbf{Conditioned dynamics on step 1---}}
The simplest setting in the conditional statistics is the conditioning on the loss of a particle in one of the two available modes $(V,-1)$ and $(H,1)$ at step 1.
In Fig.~\ref{Fig:condstep1} we present the evolution of the output distributions for both conditioning modes. The comparisons with the theory find high similarities of 96.05\% and 93.87\%, respectively. The overall similarities are evaluated according to Eq.(\ref{eq:simi}).

\begin{figure*}[t]
	\centering 
	\includegraphics[width =0.9\textwidth]{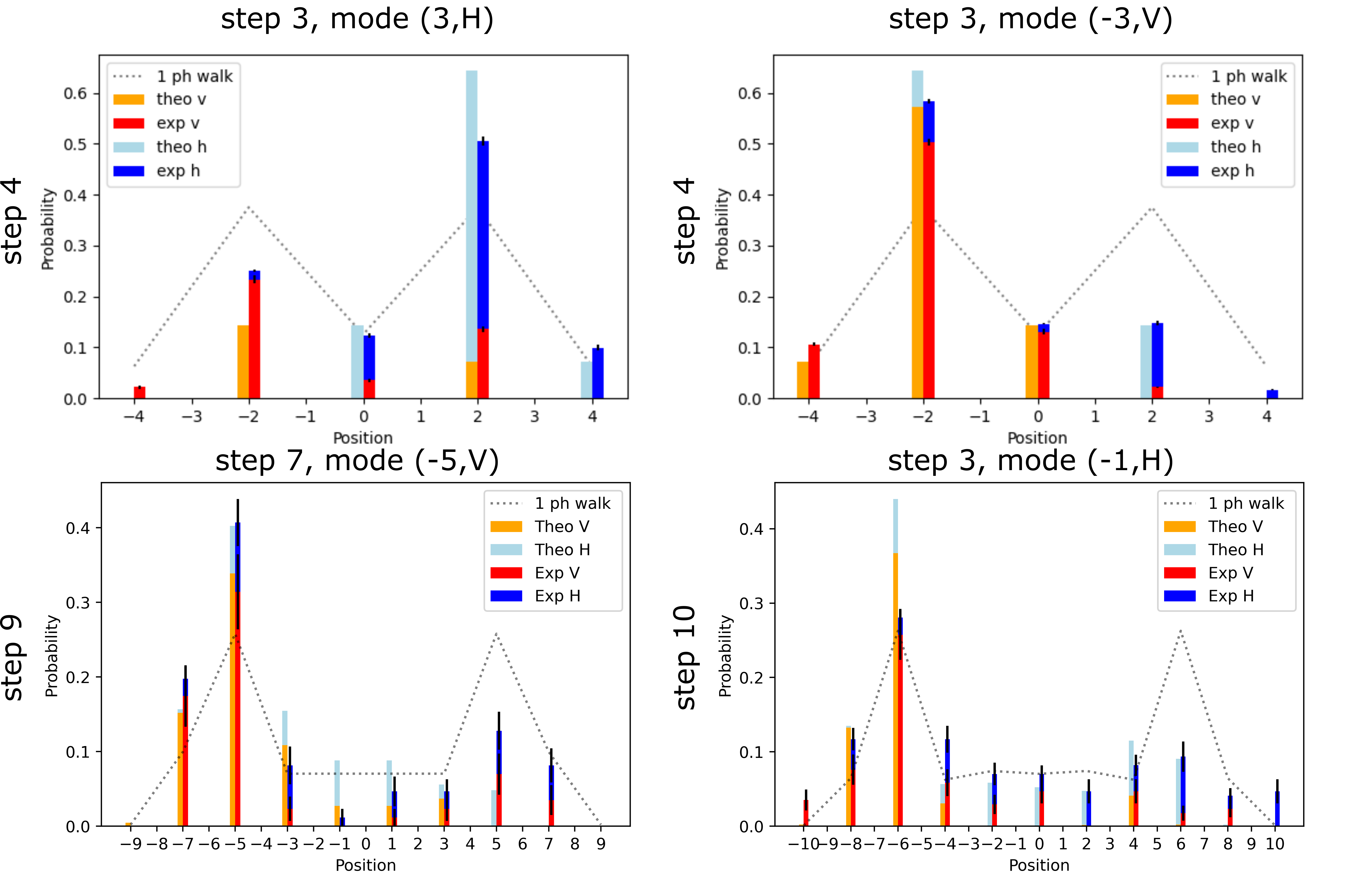}
	\caption{Dynamic conditioning for steps 4 (upper row), 9 and 10 (lower row) and different conditioning modes as indicated in the titles of the subfigures. The  similarities between experimental probability distribution and theoretical prediction  are 93.16\%  and 99.50\% (upper row), 96.54\% and 96.34\% (lower row), respectively. 
	}\label{Fig:condsteps_TM}
\end{figure*}

 The distribution of the residual photon resembles the well-known quantum walk distributions in this example. The two distributions feature a clear asymmetry and this is an effect of the initial HOM interference happening at the first step of the walk: because of it, the two photons will be in a \emph{NOON} state over the two outputs and loosing one of them in one mode will localise the other on the same mode. Therefore with this simple setting we are able to see the effects of photon bunching.
 However, the dynamic conditioning in a later step will project the residual photon onto a coherent superposition involving more modes.

 \emph{\textbf{Conditioned dynamics on steps  $>$1 ---}}
 In Fig. \ref{Fig:condsteps_TM} we present 4 probability distributions of the dynamic conditioning for various steps and conditioning  modes.
The two top ones show two output distributions that correspond to cases where one photon was lost in step 3 and the other is measured in step 4, while the bottom ones show two cases where we have lost one photon at step 7 and 3 and the surviving one has been detected at step 9 and 10, respectively. Among all the possible choices we have selected these specific plots to be shown as they provide representative examples of how conditioning impacts on the shape of the output probabilities and shows patterns that are correlated to the structure of the QW evolution network. In particular, we want to highlight the effects of mode dependent losses on the symmetry of the distributions when compared to an unperturbed case.

The upper row of Fig. \ref{Fig:condsteps_TM} shows the conditioning in step 3 in mode $(3,H)$ and $(-3,V)$, the two distributions appear highly asymmetrical and mirrored. The asymmetry of the distributions shows the effect of the conditioning, whereas the fact that the two plots appear to be mirrored is expected from the symmetry of the network, as the conditioning modes are symmetrical with respect to the initial position of the walkers.
Other two cases where the dynamic conditioning produces an asymmetrical output are shown in the bottom row of Fig. \ref{Fig:condsteps_TM} but for longer evolutions.
All the distributions exhibit large differences to the single walker one (initialised with a symmetric coin state) as indicated by the grey dotted line.

\begin{figure}[t]
	\centering 
	\includegraphics[width =0.95\columnwidth]{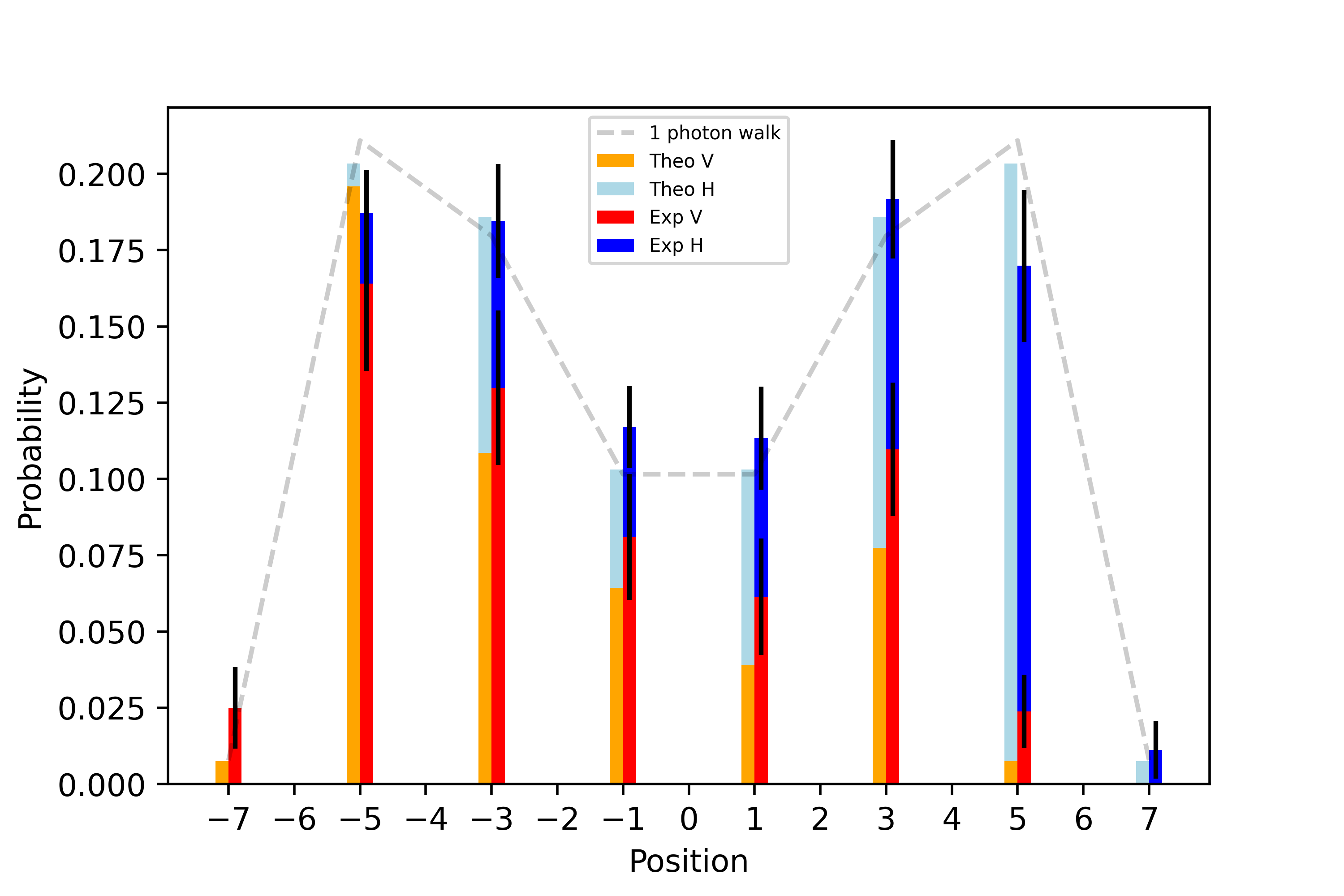}
	\caption{Dynamic conditioning in step 7 averaged over all possible loss modes from step 1 to 6. The  similarity between experimental probability distribution and theoretical prediction  is 99.41\% and 96.87\%, the dashed line in the background indicates the symmetric single photon quantum walk distribution initialised with $|\Psi\rangle =|0\rangle \otimes (|H\rangle+i|V\rangle)$ as a reference.
	}\label{Fig:average}
\end{figure}

\emph{\textbf{Conditioned dynamics averaging---}}
Here, we compare the situation of a two photon walk in which one of the photons got lost at an unknown point of the evolution with the standard one particle quantum walk. 
This means that after selecting a certain step $s$, we average the conditioned quantum walk distributions for that step on all the loss configurations ranging from step 1 to $s-1$.  
Thereby we pretend, that we do not have any knowledge about the intermediate state of the walk. In Fig.~\ref{Fig:average} we present the average distribution obtained at step 7, we immediately observe that it is impossible to experimentally distinguish the two settings in strong contrast to the settings of Fig.~\ref{Fig:condsteps_TM}, where the conditioned distribution is highly different from that of a single photon walk.

\begin{figure}[t]
	\centering 
	\includegraphics[width =0.95\columnwidth]{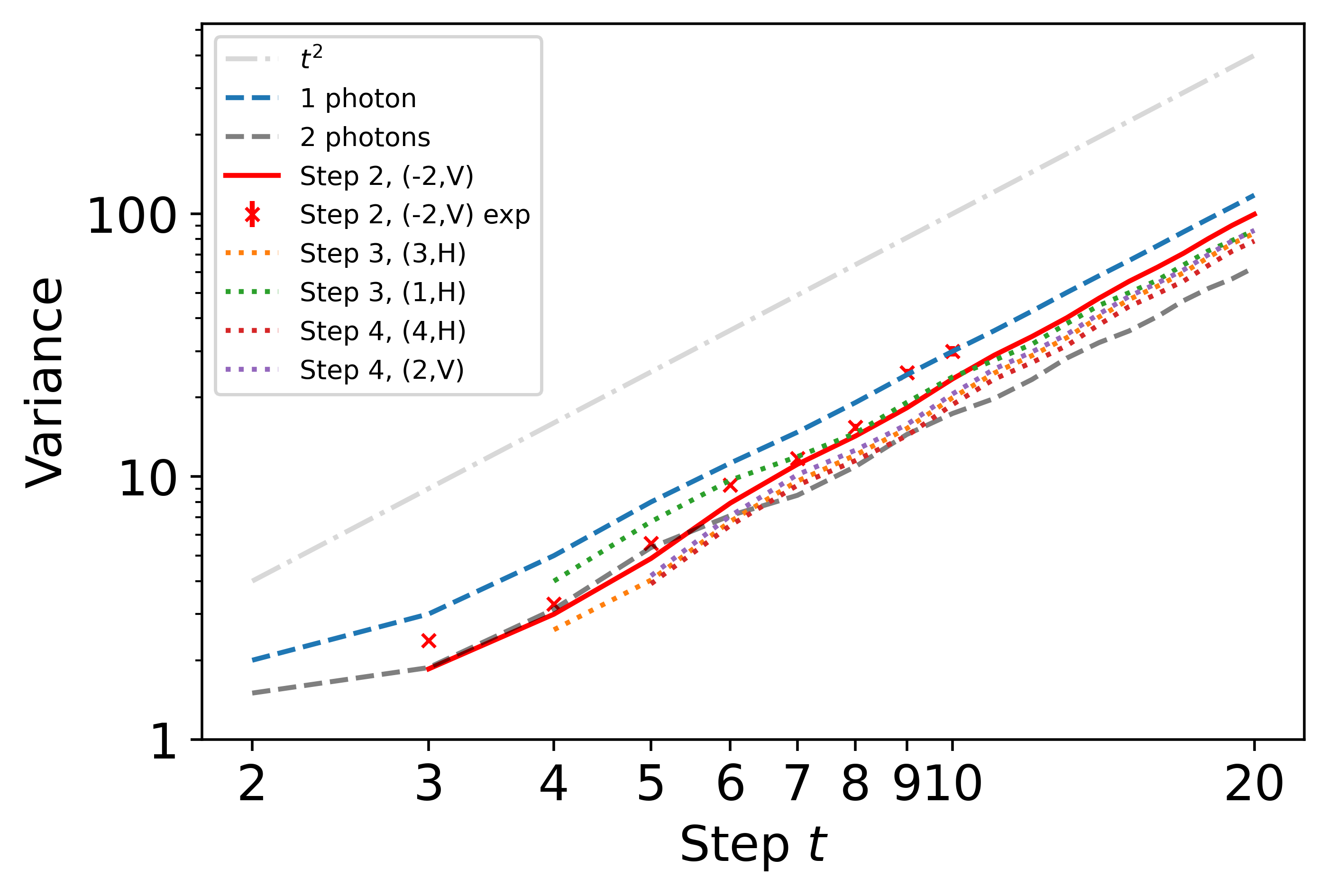}
	\caption{Asymptotic spreading of one photon walk (blue), 2 photon quantum walk (grey), and conditioned walks (dotted lines). The red crosses indicate one exemplary set of experimental data for the conditioning in step 2 and mode $(-2,V)$ in accordance with the theory data (solid red). The grey dotted-dashed line shows a constant slope of 2, the expected ballistic spread. The data are reported together with the estimated errors which are too small to be visible in the plot 
	}\label{Fig:asymptotic}
\end{figure}

\emph{\textbf{Asymptotic scaling---}}
Now we study if the dynamic conditioning changes the asymptotic scaling of the one particle dynamics, i.e. the ballistic spread $\mathrm{Var}(t) \propto t^2$.
In Fig.~\ref{Fig:asymptotic} we plot the variance of the distribution of some examples of conditioned dynamics and compare it with the spread of the single particle walk and the two photon walk.
We determine the variances as
\begin{align}
\mathrm{Var}_{1d} &= \sum_{x}x^2P(x)-\left( \sum_{x}xP(x)\right)^2\\
\mathrm{Var}_{2d} &= \sum_{(x,y)}\left(\frac{x+y}{2}\right)^2P(x,y)-\left( \sum_{(x,y)}\frac{x+y}{2}P(x,y)\right)^2, \nonumber  
\end{align}
where $x$ and $y$ are the quantum walk positions.
The dotted-dashed light gray line indicates the ballistic spread with a constant slope of 2.

One observes clearly, that after the transient behaviour in the first few steps all distributions spread ballistically with the same slope.
Thus, a dynamic conditioning does not impact the spreading characteristics of the asymptotic distribution, but just modifies the transient behaviour within the first steps. In the same picture we show a set of experimental points (red crosses) obtained employing the TM setup. They show the measured variance trend for the output probability distributions conditioned on losses at step 2 and mode $(-2,V)$ and output steps from 2 to 10, the data are reported together with the estimated errors which are too small to be visible in the plot.

\section{Conclusion and Outlook}
\label{Sec:conclusion}

\begin{figure}[t]
	\centering 
	\includegraphics[width =0.9\columnwidth]{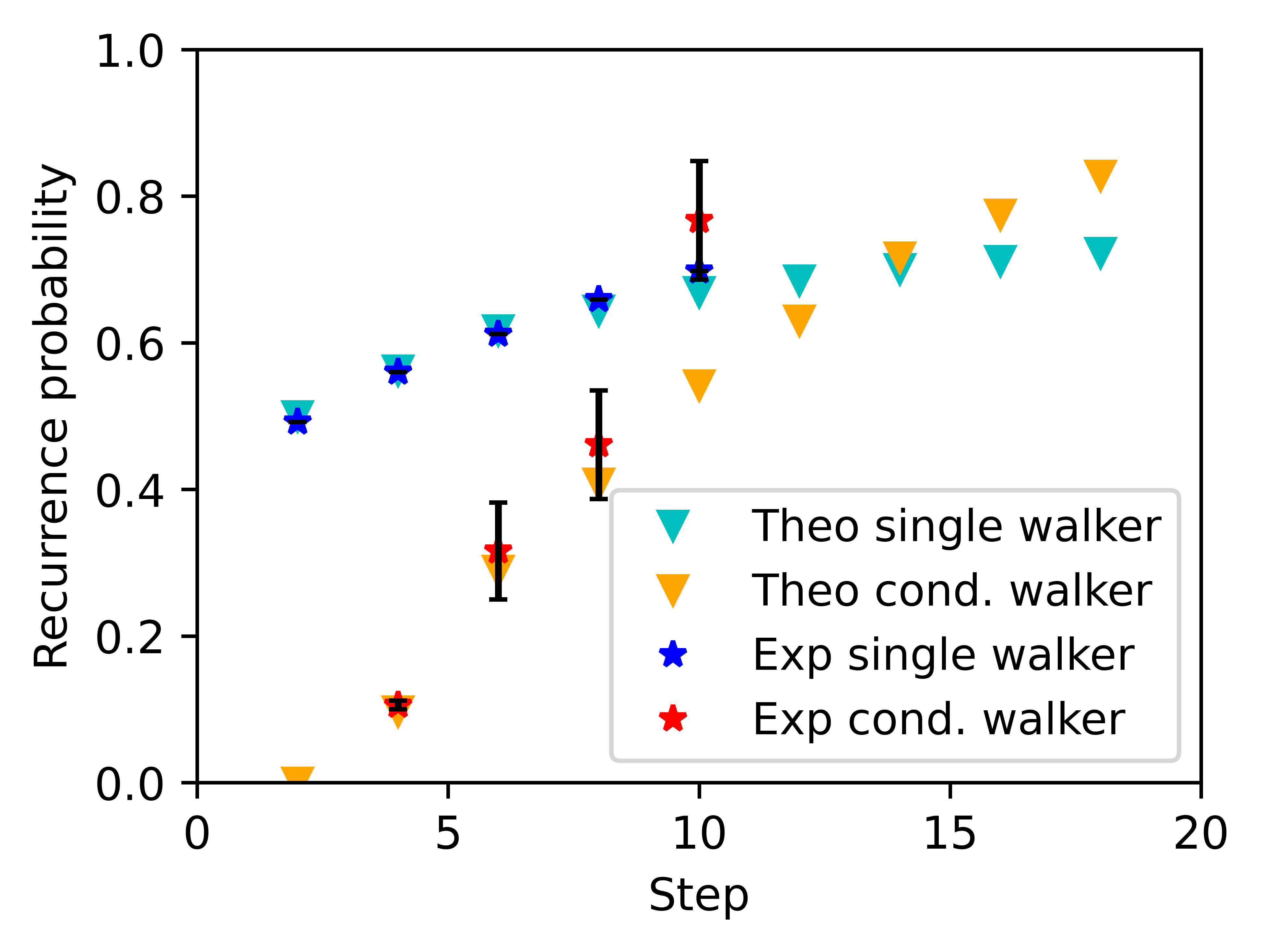}
	\caption{Simulated and measured trends of the return probability to $x = 0$ within the first $T$ steps of a single walk (turquoise) and a conditioned walker given that the first walker was there before in any step $t<T$ (orange symbols).
	}\label{Fig:recurrence}
\end{figure}

In this work, we studied the basic principle of conditioned dynamics in two photon DTQWs. We have developed the theoretical formalism to include particle losses in the QW evolution and we have shown an experimental setup capable of implementing this formalism employing TM.
We observe that, knowing where one of the two photons was lost, the output distribution shows pronounced differences with respect to the unperturbed case.
We find that averaging over all possible loss modes destroys the effect of the dynamic conditioning and the final distribution cannot be experimentally distinguished from a single particle walk anymore.
When focusing on the asymptotic behaviour, our simulations show that the conditioning does not influence the ballistic spread of the walker and that after a short transient behaviour the variances of the spatial probability distributions grow with the square of the step number.

Our work will lay the foundation to address a variety of quantum passage and meeting problems with multiple quantum agents \cite{stefanak2006meeting}.
One example would be a (quantum) civilisation problem, that addresses the question how likely it is that a second walker visits a particular position given that the first walker has explored that terrain already some time before.
This question is a multi-walker generalisation of the recurrence problem \cite{ stefanak2008recurrence, stefanak_recurrence_2008,nitsche2018probing}.
The numerical simulation of this probability shows, that after the first few steps the conditioned recurrence to the initial position exceeds the recurrence of an independent single walker and approaches 1 (i.e. returns with certainty) much faster, Fig.~\ref{Fig:recurrence}. In the same picture, we show a set of experimental data obtained using the TM loop for a single and two photon 10 step quantum walk (blue and red stars). In general, data and theoretical predictions follow the expected trend, with some discrepancies  for large step numbers due to the accumulation of experimental inaccuracies.

In addition, the setting of dynamic conditioning serves as a discrete variable variant of photon-subtraction protocols which are established in CV field to increase and distil entanglement between CV states \cite{biswas2007nonclassicality,dodonov2009smooth,ourjoumtsev2007increasing,parigi2007probing,lee2012generating,averchenko2014nonlinear,ra2017tomography}. 

In continuous-variable context photon subtraction protocols were used to generate non-Gaussian quantum states of light \cite{walschaers2018tailoring,ra2020non}.
Transformed into the  discrete variable formalism, our results can pave a way to generate complex quantum superposition states with multiple parties.

\begin{acknowledgments}
We acknowledge financial support by the European Research Council through the ERC project QuPoPCoRN (Grant No. 725366).
\end{acknowledgments}



\bibliography{thispaper_bib}
\end{document}